\documentclass[titlepage,oneside,12pt]{article}

\usepackage[tiny,rm]{titlesec}
\newpagestyle{trbstyle}{
	\sethead{Bandini S., Crociani L. and Vizzari G.}{}{\thepage}
}
\titleformat{\section}{\bfseries}{}{0pt}{\uppercase}
\titleformat{\section}{\bfseries}{\arabic{section}~}{0pt}{\uppercase}
\titleformat{\subsection}{\bfseries}{\arabic{section}.\arabic{subsection}~}{0pt}{}
\titlespacing*{\subsection}{0pt}{12pt}{*0}
\titleformat{\subsubsection}{\itshape}{\arabic{section}.\arabic{subsection}.\arabic{subsubsection}~}{0pt}{}
\titlespacing*{\subsubsection}{0pt}{12pt}{*0}

\usepackage{enumitem}
\setlist[1]{labelindent=0.5in,leftmargin=*}
\setlist[2]{labelindent=0in,leftmargin=*}

\usepackage{ccaption}
\usepackage{amsmath}
\makeatletter
\renewcommand{\fnum@figure}{\textbf{FIGURE~\thefigure} }
\renewcommand{\fnum@table}{\textbf{TABLE~\thetable} }
\makeatother
\captiontitlefont{\bfseries \boldmath}
\captiondelim{\;}
\usepackage[sort,numbers]{natbib}
	
\setcitestyle{round}

\usepackage[pagewise]{lineno}

\usepackage{totcount}
	\regtotcounter{table} 	%count tables
	\regtotcounter{figure} 	%count figures
	\regtotcounter{algorithm} %count algorithms

\newcommand\wordcount{
    \immediate\write18{texcount -sum -1 \jobname.tex > 'count.txt'} 5739
 }

\usepackage{mathptmx}

\usepackage[T1]{fontenc}
\usepackage{textcomp}
\usepackage[noae]{Sweave}

\usepackage{graphicx}
\usepackage{subfigure}
\usepackage{algorithm}
\usepackage{algpseudocode}
\usepackage{amsmath}
\usepackage{amssymb}
\usepackage{url}

\begin{document}

	\pagewiselinenumbers % comment out for final manuscript
	\thispagestyle{empty}

\begin{titlepage}
\begin{flushleft}

% Title
{\LARGE \bfseries Heterogeneous Speed Profiles in Discrete Models for Pedestrian Simulation}\\[1cm]

Stefania Bandini\\
CSAI - Complex Systems \& Artificial Intelligence Research Center\\
University of Milano-Bicocca\\
Viale Sarca 336/14\\
20126, Milano, Italy\\
tel. +390264487835, fax. +390264487839\\
bandini@disco.unimib.it\\[1cm]

Luca Crociani\\
CSAI - Complex Systems \& Artificial Intelligence Research Center\\
University of Milano-Bicocca\\
Viale Sarca 336/14\\
20126, Milano, Italy\\
tel. +390264487857, fax. +390264487839\\
luca.crociani@disco.unimib.it\\[1cm]

Giuseppe Vizzari*\\
CSAI - Complex Systems \& Artificial Intelligence Research Center\\
University of Milano-Bicocca\\
Viale Sarca 336/14\\
20126, Milano, Italy\\
tel. +390264487865, fax. +390264487839\\
vizzari@disco.unimib.it\\[1cm]

* corresponding author\\[1cm]

\begin{center}
Submitted for presentation and publication\\
Transportation Research Record\\
Committee number AHB45\\
TRB Committee on Traffic Flow Theory and Characteristics\\[1cm]

\wordcount words + \total{figure} figures + \total{table} tables + \total{algorithm} algorithms @ 250 = 7489 (limit 7500)

\today

\end{center}
\end{flushleft}
\end{titlepage}

\newpage

\thispagestyle{empty}
\section*{Abstract}

Discrete pedestrian simulation models are viable alternatives to particle based approaches based on a continuous spatial representation. The effects of discretisation, however, also imply some difficulties in modelling certain phenomena that can be observed in reality. This paper focuses on the possibility to manage heterogeneity in the walking speed of the simulated population of pedestrians by modifying an existing multi-agent model extending the floor field approach. Whereas some discrete models allow pedestrians (or cars, when applied to traffic modelling) to move more than a single cell per time step, the present work proposes a maximum speed of one cell per step, but we model lower speeds by having pedestrians yielding their movement in some turns. Different classes of pedestrians are associated to different desired walking speeds and we define a stochastic mechanism ensuring that they maintain an average speed close to this threshold. In the paper we formally describe the model and we show the results of its application in benchmark scenarios. Finally, we also show how this approach can also support the definition of slopes and stairs as elements reducing the walking speed of pedestrians climbing them in a simulated scenario.

\newpage

\section{Introduction}

The modelling and simulation of pedestrians and crowds is a consolidated and successful application of research results in the more general area of computer simulation of complex systems. It is an intrinsically interdisciplinary effort, with relevant contributions from disciplines ranging from physics and applied mathematics to computer science, often influenced by (and sometimes in collaboration with) anthropological, psychological, sociological studies and the humanities in general. The level of maturity of these approaches was sufficient to lead to the design and development of commercial software packages, offering useful and advanced functionalities to the end user (e.g. CAD integration, CAD-like functionalities, advanced visualisation and analysis tools) in addition to a simulation engine (see ~\url{http://www.evacmod.net/?q=node/5} for a large although not necessarily complete list of pedestrian simulation models and tools). Nonetheless, as testified by a recent survey of the field by~\cite{DBLP:reference/complexity/SchadschneiderKKKRS09} and by a report commissioned by the Cabinet Office by~\cite{UnderstandingCrowd}, there is still much room for innovations in models improving their performances both in terms of \emph{effectiveness} in modelling pedestrians and crowd phenomena, in terms of \emph{expressiveness} of the models (i.e. simplifying the modelling activity or introducing the possibility of representing phenomena that were still not considered by existing approaches), and in terms of \emph{efficiency} of the simulation tools. Research on models able to represent and manage phenomena still not considered or properly managed is thus still lively and important.

Discrete pedestrian simulation models are viable alternatives to particle based models, based on a continuous spatial representation (see, e.g.,~\cite{DBLP:reference/complexity/SchadschneiderKKKRS09}) and they are able to reproduce realistic pedestrian dynamics from the point of view of a number of observable properties. The effects of discretisation, however, also imply some difficulties in modelling certain phenomena that can be observed in reality. This paper focuses on the possibility of modelling heterogeneity in the walking speed of the simulated population of pedestrians, a characteristic reported and analysed, for instance, in~\cite{Willis2004} (as observed in common streets in York and Edimburgh) or~\cite{Schultz2008} (as observed in the Dresden Airport). In particular, this paper proposed a relevant modification of an existing multi-agent model extending the floor field approach~\cite{CASM2013}. Relevant previous works describe discrete models allowing pedestrians to move more than a single cell per time step~\cite{klupfel2003,DBLP:conf/acri/Yamamoto08,DBLP:conf/acri/KretzKS08}, an approach that however requires the introduction of specific rules to manage unavoidable additional conflicts that arise from the potential crossings of pedestrians' paths. In the present work, instead, we maintain a maximum speed of one cell per step for each pedestrian, but we model lower speeds by having pedestrians yielding their movement in some turns (an approach that was adopted in~\cite{Shimura2013} for modelling elderly persons). Different classes of pedestrians are associated to different desired walking speeds and we define a stochastic mechanism ensuring that they maintain an average speed close to this threshold. In the paper we will formally describe the model and we will show the results of its application in benchmark scenarios (single and counter flows in simple scenarios). Finally, we will also show how this approach can also support the definition of slopes and stairs as elements reducing the walking speed of pedestrians climbing them in a simulated scenario.

\section{A Discrete Model for Pedestrian Simulation}

This section provides a formal description and discussion of a discrete and agent--based computational model for crowd simulation in normal conditions (evacuation scenarios are currently not considered). The main innovative feature of the model is the fact that pedestrian \emph{groups} are considered as a significant element that can influence the dynamics of the overall system. The model has been developed in the \textbf{MakkSim} platform~\cite{CrocianiMV13}, a pedestrian dynamics simulator built as a plug--in for \emph{Blender} software (\url{http://www.blender.org}), and it will be described through its main elements.

\subsection{Environment}

The environment is modeled in a discrete way by representing it as a grid of squared cells with $40$ $cm^2$ size (according to the average area occupied by a pedestrian~\cite{weidmann}). Cells have a state indicating the fact that they are vacant or occupied by obstacles or pedestrians: 
\begin{center}
$State(c) : Cells \rightarrow \left\lbrace Free,\ Obstacle,\ OnePed_i, TwoPeds_{ij} \right\rbrace$
\end{center}
The last two elements of the definition point out if the cell is occupied by one or two pedestrians respectively, with their own identifier: the second case is allowed only in a controlled way to simulate overcrowded situations, in which the density is higher than $6.25$ $m{^2}$ (i.e. the maximum density reachable by our discretisation).

A simulation scenario includes both the structure of the environment and all the information required for the execution of a specific simulation, such as crowd management demands (pedestrians generation profile, origin-destination matrices) and spatial constraints. Some relevant entities of a scenario are represented in the model in terms of \emph{spatial markers}, special sets of cells that describe relevant elements in the environment. In particular, three kinds of spatial markers are defined:

\begin{itemize}
\item $start$ areas, that indicate the generation points of agents in the scenario. Agent generation can occur in \textit{block}, all at once, or according to a user defined \textit{frequency}, along with information on type of agent to be generated and its destination and group membership;
\item $destination$ areas, which define the possible targets of the pedestrians in the environment;
\item $obstacles$, that identify all the areas where pedestrians cannot walk.
\end{itemize}

Space annotation allows the definition of virtual grids of the environment, as containers of information for agents and their movement. In our model, we adopt the \textit{floor field} approach \cite{Burstedde2001}, that is based on the generation of a set of superimposed grids (similar to the grid of the environment) starting from the information derived from spatial markers. Floor field values are spread on the grid as a gradient and they are used to support pedestrians in the navigation of the environment, representing their interactions with static objects (i.e., destination areas and obstacles) or with other pedestrians. Moreover, floor fields can be \emph{static} (created at the beginning and not changed during the simulation) or \emph{dynamic} (updated during the simulation). Three kinds of floor fields are defined in our model:
\begin{itemize}
\item \textit{path field}, that indicates for every cell the distance from one destination area, acting as a potential field that drives pedestrians towards it (static). One path field for each destination point is generated in each scenario;
\item \textit{obstacles field}, that indicates for every cell the distance from neighbor obstacles or walls (static). Only one obstacles field is generated in each simulation scenario;
\item \textit{density field}, that indicates for each cell the pedestrian density in the surroundings at the current time-step (dynamic). Like the previous one, the density field is unique for each scenario.
\end{itemize}

Chessboard metric with $\sqrt{2}$ variation over corners \cite{Kretz2010} is used to produce the spreading of the information in the path and obstacle fields. Moreover, pedestrians cause a modification to the density field by adding a value $v = \frac{1}{d^2}$ to cells whose distance $d$ from their current position is below a given threshold $R$. Since $\lim_{r \rightarrow 0^+}\frac{1}{d^2}=\infty$, this calculation is applied only for the cells of the neighbourhood, excluding the cell in which the agent is situated, for which the value 1 is added to the density field. In this way, values contained in each cell are 4 times the local density measured in a circle of radius $R$.

Agents are able to perceive floor fields values in their neighborhood by means of a function $Val(f,c)$ ($f$ represents the field type and $c$ is the perceived cell). This approach to the definition of a perception model moves the burden of its management from agents to the environment, which would need to monitor agents anyway in order to produce some of the simulation results.

\subsection{Pedestrians and Movement}
\label{sec:agent}

Formally, our agents are defined by the following triple:
\begin{center}
$Ped : \left\langle Id,\ Group,\ State \right\rangle$;\ \ \  $State : \left\langle position,\ oldDir,\ Dest \right\rangle$
\end{center}
with their own numerical identifier, their group (if any) and their internal state, that defines the current position of the agent, the previous movement and the final destination, associated to the relative path field.

Before describing agent behavioural specification, it is necessary to introduce the formal representation of the nature and structure of the groups they can belong to, since this is an influential factor for movement decisions. 

\subsubsection{Social Interactions}
The model aims at representing social relationships that we assume to be stable in the simulated scenario. Two kinds of groups have been defined: the \textit{simple group}, that indicates a family, a small set of friends, or any other small group in which there is a strong and simply recognisable cohesion; the \textit{structured group}, generally a large one (e.g. a group of team supporters or a touristic group), that shows a slight cohesion and a natural fragmentation into subgroups. 

Between members of a simple group like a family it is possible to identify an apparent tendency to stay quite close, in order to guarantee the possibility to perform interactions by means of verbal or non-verbal communication~\cite{CostaGroupDistances}. On the contrary, in large groups people are mostly linked by the sharing of a common goal, and the overall group tends to maintain only a week compactness, with a following behaviour between members. In order to model these two typologies, the formal representation of a group is described by the following:
\begin{center}
$Group : \left\langle Id,\left[SubGroup_1,\dots,SubGroup_m\right],\left[Ped_1,\cdots,Ped_n\right]\right\rangle$
\end{center}
In particular, if the group is simple, it will have an empty set of subgroups, otherwise it will not contain any direct references to pedestrians inside it, which will be stored in the respective leafs of its tree structure. Differences on the modelled behavioural mechanism in simple/structured groups will be analysed in the following section, with the description of the utility function.

\subsubsection{Agent Behavior}
\label{sec:agentBehaviour}
In order to perform the agent behaviour, its life-cycle has been defined on four steps: \emph{perception, utility calculation, action choice} and \emph{movement}. The \textit{perception} step provides to the agent all the information needed for choosing its destination cell. In particular, if an agent does not belong to a group (from here called \textit{individual}), in this phase it will only extract values from the floor fields, while in the other case it will perceive also the positions of the other group members within a configurable distance, for the calculation of the overall \textit{cohesion} parameter. The choice of each action is based on an utility value, that is assigned to every possible movement according to the following function:
\begin{center}
$U(c)=\frac{\kappa_g G(c)+\kappa_{ob} Ob(c)+\kappa_s S(c)+\kappa_{c} C(c)+\kappa_{i} I(c)+\kappa_{d} D(c)+\kappa_{ov} Ov(c)}{d}$
\end{center}
Function $U(c)$ takes into account the behavioural components considered relevant for pedestrian movement. Each one is modelled with a function that returns values in range $[-1;+1]$, if it represents an \textit{attractive} element (negative values are associated to costs associated, for instance, to movements leading father from the desired destination), or in range $[-1;0]$, if it represents a \textit{repulsive} one for the agent (in this case, the lack of a repulsive element does not represent a benefit). For each function has been introduced a $\kappa$ coefficient for its calibration. It must be noted that the idea of attractive or repulsive forces guiding the behaviour of pedestrians mainly comes from particle based models, whose main representative is the \emph{social force} model~\cite{HelbingSocialForce}, encapsulating basic proxemic~\cite{Hall-HiddenDimension} considerations into a dynamical model.

The first three functions combine information derived by local floor fields and they model the basic factors considered in the pedestrian behaviour: goal attraction (i), geometric (ii) and social repulsion (iii). The fourth and fifth elements aggregate the perceived positions of members of agent group, both simple (iv) and structured (v), to calculate the level of attractiveness of each neighbour cell, relating to cohesion phenomenon. Moreover, two factors represent preferences with respect to movement, helping the model to reproduce more realistic simulations both in qualitative and quantitative perspective: (vi) adds a bonus to the utility of the cell next to the agent according to his/her previous direction, while (vii) describes the \textit{overlapping} mechanism, a method used to allow our model the possibility to treat high density situations, allowing two pedestrians temporarily occupying the same cell at the same step.

The purpose of the function denominator $d$ is to provide a penalisation to the diagonal movements, since their selection is actually associated to a movement covering a higher distance than the one associated to the other directions. As shown later, the method for heterogeneous speed profiles described in this paper also considers this issue. 

\begin{figure}[tb]
\centering
\includegraphics[width=\columnwidth] {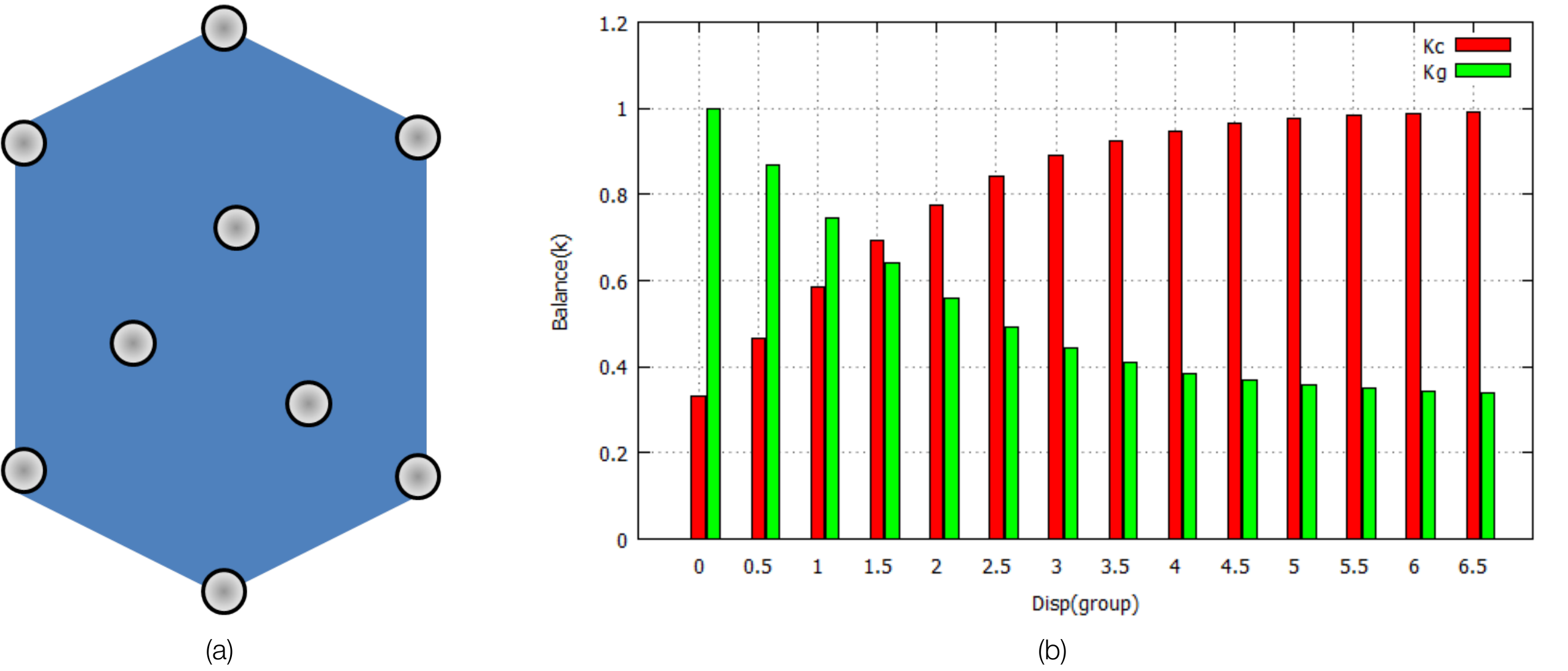}
\caption{Graphical representation of a group composed of nine members and the area of the convex polygon that contains all the group members (a) and graph of $Balance(k)$, for $k=1$ and $\delta = 2.5$ (b).}
\label{fig:area_group}
\end{figure}

As we previously mentioned, the main difference between simple and structured groups is associated to the cohesion intensity, that is much higher for simple groups. Functions $C(c)$ and $I(c)$ have been defined to correctly model this difference. Nonetheless, various preliminary tests on benchmark scenarios show us that, used singularly, function $C(c)$ is not able to reproduce realistic simulations. Human behaviour is, in fact, very complex and can react differently even in simple situation, for example by allowing temporary fragmentation of simple groups in front of several constraints (obstacles or opposite flows). Acting statically on the calibration weight, it is not possible to configure this dynamic behaviour: with a small cohesion parameter several permanent fragmentations have been reproduced, while with an increase of it we obtained no group dispersion, but also an excessive and unrealistic compactness of them.

In order to face this issue, another function has been introduced in the model, with the aim to balance the calibration weight of the three attractive behavioural elements, depending from the fragmentation level of simple groups:
\begin{center}
$Balance(k) =
\begin{cases}
\frac{1}{3} \cdot k + (\frac{2}{3} \cdot k \cdot DispBalance) & \textit{if } k = k_c \\
\frac{1}{3} \cdot k + (\frac{2}{3} \cdot k \cdot (1 - DispBalance)) & \textit{if } k = k_g \vee k=k_i\\
k & \textit{otherwise}
\end{cases}$
\end{center}
\begin{center}
$DispBalance =  tanh \left(  \displaystyle \frac{Disp(Group)}{\delta} \right);  \ \ \ Disp(Group)=\frac{Area(Group)}{|Group|}$
\end{center}
where $k_i$, $k_g$ and $k_c$ are the weighted parameters of $U(c)$, $\delta$ is the calibration parameter of this mechanism and $Area(Group)$ calculates the area of the convex hull defined using positions of the group members. Fig.~\ref{fig:area_group} exemplifies both the group dispersion computation and the effects of the $Balance$ function on parameters. A dynamic and adaptive behaviour of groups has been obtained with this mechanism, which relaxes the cohesion if members are sufficiently compact and intensifies it with the growing of dispersion.

After the utility evaluation for all the cells in the neighbourhood, the choice of action is  stochastic, with the probability to move in each cell $c$ as ($N$ is the normalisation factor):
\begin{center}
$P(c) = N \cdot e^{U(c)}$
\end{center}
On the basis of $P(c)$, agents move in the resulted cell according to their set of possible actions, defined as list of the eight possible movements in the Moore neighbourhood, plus the action to keep the position (indicated as $X$): $A = \{NW,N,NE,W,X,E,SW,S,SE\}$.

\subsection{Time and Update Mechanism}

In the basic model definition time is also discrete; in an initial definition of the duration of a time step was set to 1/3 of second. This choice, considering the size of the cell (a square with 40 cm sides), generates a linear pedestrian speed of about $1.2$ $ms^{-1}$, which is in line with the data from the literature representing observations of crowd in normal conditions \cite{weidmann}.

Regarding the update mechanism, three different strategies are usually considered in this context~\cite{klupfel2003}: \textit{ordered sequential}, \textit{shuffled sequential} and \textit{parallel} update. The first two strategies are based on a sequential update of agents, respectively managed according to a \textit{static} list of priorities that reflects their order of generation or a \textit{dynamic} one, shuffled at each time step. On the contrary, the parallel update calculates the choice of movement of all the pedestrians at the same time, actuating choices and managing conflicts in a latter stage. The two sequential strategies, instead, imply a simpler operational management, due to an a-priori resolution of conflicts between pedestrians. In the model, we adopted the parallel update strategy. This choice is in accordance with the current literature, where it is considered much more realistic due to consideration of conflicts between pedestrians, arisen for the movement in a shared space~\cite{KirchnerPED2003,Nishinari2003}.

With this update strategy, the agents life-cycle must consider that before carrying out the \textit{movement} execution potential conflicts, essentially related to the simultaneous choice of two (or more) pedestrians to occupy the same cell, must be solved. The overall simulation step therefore follows a three step procedure:

\begin{itemize}
\item \emph{update of choices} and \emph{conflicts detection} for each agent of the simulation;
\item \textit{conflicts resolution}, that is the resolution of the detected conflicts between agent intentions;
\item \textit{agents movement}, that is the update of agent positions exploiting the previous conflicts resolution, and \textit{field update}, that is the computation of the new density field according to the updated positions of the agents.
\end{itemize}

The resolution of conflicts employs an approach essentially based on the one introduced in~\cite{Nishinari2003}, based on the notion of friction. Let us first consider that conflicts can involve two of more pedestrians: in case more than two pedestrians involved in a conflict for the same cell, the first step of the management strategy is to block all but two of them, chosen randomly, reducing the problem to the case of a simple conflict among two pedestrians. To manage a simple conflict, another random number between 0 and 1 is generated and compared to two thresholds, $\mathit{frict}_l$ and $\mathit{frict}_h$, with $0 < \mathit{frict}_l < \mathit{frict}_h \leq 1$: the outcome can be that all agents are blocked when the extracted number is lower than $\mathit{frict}_l$, only one agent moves (chosen randomly) when the extracted number is between $\mathit{frict}_l$ and $\mathit{frict}_h$ included, or even two agents move when the number is higher than $\mathit{frict}_h$ (in this case pedestrian overlapping occurs). For our tests, the values of the thresholds make it quite relatively unlikely the resolution of a simple conflict with one agent moving and the other blocked, and much less likely their overlapping.

\section{The Heterogeneous Speed Profile Extension}
In the current literature, discrete models generally assume only one speed profile for all the population and this is considered one of the main limitations of this approach. With the extension described in this section we want to overcome this limitation in an efficient and realistic way, that is still able to reproduce the overall dynamics.

Firstly, an exploration of the relevant aproaches to our problem is needed. Modelling different speed profiles in discrete models can be done in two principal ways (the first one has been introduced in \cite{Kirchner2004})

%\begin{figure} [t!]
%\begin{center}
%\includegraphics[width=.5\columnwidth]{Images/conflicts.pdf}
%\end{center}
%\caption{Examples of possible conflicts arising when velocities above one cell per turn are allowed in discrete models: in addition to already existing possible conflicts on the destination of a pedestrian movement, even potentially illegal crossing paths must be considered, effectively requiring the modeling of sub-turns.}
%\label{fig:vMax}
%\end{figure}

\begin{enumerate}
\item By increasing agents movement capabilities (i.e. they can move more than 1 cell per time step), according to their \emph{desired speed}. In this way, given $k$ the side of cells of the discrete grid, its possible to obtain speed profiles equal to $n \cdot k$ m/step, with $n \in \mathbb{N}$ equal to the maximum number of movements per step.
%\item By improving movement capabilities of the agents and modifying the current space discretization towards a finer grain, described by a lower size of cells. Thanks to this method, by fixing a maximum speed of pedestrians, it is possible to obtain a greater set of possible speed profiles.
\item By modifying the current time scale, making it possible to cover the same distance in less time and achieving thus a higher maximum speed profile but at the same time allowing each pedestrian to yield their turn in a \textit{stochastic} way according to an individual parameter, achieving thus a potentially lower speed profile.
%\item By refining the space discretization and also the time-scale, leaving agents movement capabilities located in the cells surrounding the ones where they are located. As for the previous one, also in this way the desired speed of each pedestrian is obtained in a stochastic way.
\end{enumerate}

Naturally, both methods can be more effective with a finer grained discretisation, which supports a more precise representation of the environment and the micro-interactions between pedestrian~\cite{Kirchner2004}, but the simulation would be less efficient. Computational costs increase proportionally to the ratio $S_o / S_n$, with $S_o \mbox{ and } S_n$ respectively equal to the old and new size of cells (e.g. if the size is halved, for performing the same space agents will need a number of update cycles at least doubled).

The method supporting movements of more than a single cell can be effective, but it leads to complications and increased computational costs for the managing of micro-interactions and conflicts: in addition to already existing possible conflicts on the destination of two (or more) pedestrian movements, even potentially illegal crossing paths must be considered, effectively requiring the modelling of sub-turns. Therefore, we decided to retain a maximum velocity of one cell per turn, but shortening the turn duration and introducing a stochastic yielding for representing speed profiles lower than the maximum.

%Also the fourth one suffers for the increasing of computational times of the simulations, which increases proportionally to the ratio $S_o / S_n$, with $S_o \mbox{ and } S_n$ respectively equal to the old and new size of cells (e.g. if the size is half-divided, for performing the same space agents will need a number of update cycles at least doubled). These reasons led to undertake the third method, since it does not affect computational complexity in a relevant way and it is the simplest one to develop.

The computational model has therefore been modified in several parts. Each agent has a new parameter $Speed_d$ in its $State$, describing its desired speed. For the overall scenario, a parameter $Speed_m$ is introduced for indicating the maximum speed allowed during the simulation (described by the assumed time scale). In order to obtain the desired speed of each pedestrian during the simulation, the agent life-cycle is then \emph{activated} according to the probability to move at a given step $\rho = \frac{Speed_d}{Speed_m}$. 

By using this method, the speed profile of each pedestrian is modelled in a fully stochastic way and, given a sufficiently high number of step, their effective speed will be equal to the wanted one. But it must be noted that in a several cases the speed has to be rendered in a relatively small time and space window (think about speed decreasing on a relatively short section of \emph{stairs}). 

In order to overcome this issue, we decided to consider $\rho$ as an indicator to be used to decide if an agent can move according to an \textit{extraction without replacement} principle. For instance, given $Speed_d=1.0 m/s$ of an arbitrary agent and $Speed_m=1.6 m/s$, $\rho$ is associated to the fraction $5/8$, that can be interpreted as an \textbf{urn model} with 5 \emph{move} and 3 \emph{do not move} events. At each step, the agent extracts once event from its urn and, depending on the result, it moves or stands still. The extraction is initialised anew when all the events are extracted. The mechanism can be formalised as follows:

\begin{algorithm} [!t]
\caption{Life-cycle update with heterogeneous speed}
\label{alg:HetUpdate}
\begin{algorithmic}
\If {$Random() \leq \alpha / \beta$}
	\If {$updatePosition()$ == true}
		\State $\alpha \gets \alpha - 1$
	\Else
		\State $\beta \gets \beta +1$
	\EndIf
\EndIf
\State $\beta \gets \beta -1$
\If{$\beta == 0$}
	\State $(\alpha,\beta) = Frac(\rho)$
\EndIf
\end{algorithmic}
\end{algorithm}

\begin{itemize}
\item Let $Frac(r): \mathbb{R} \rightarrow \mathbb{N}^2$ be a function which returns the minimal pair $(i,j) : \frac{i}{j} = r$.
\item Let $Random$ be a pseudo-random number generator.
\item Given $\rho$ the probability to activate the life-cycle of an arbitrary agent, according to its own desired speed and the maximum speed configured for the simulation scenario. Given $(\alpha,\beta)$ be the result of $Frac(\rho)$, the update procedure for each agent is described by the pseudo-code of Alg.~\ref{alg:HetUpdate}. The method $updatePosition()$ describes the attempt of movement by the agent: in case of failure (because of a conflict), the urn is not updated.
\end{itemize}

This basic mechanism allows the synchronisation between the effective speed of an arbitrary agent and its desired one every $\tau$ steps, with $\tau$ limited to $Speed_m \cdot 10^\iota$ step and $\iota$ is associated to the maximum number of decimal positions between $Speed_d$ and $Speed_m$. For instance, if the desired speed is fixed at $1.3 m/s$ and the maximum one at $2.0 m/s$, the resulting $Frac(\rho)=\frac{13}{20}$, therefore the agent average velocity will match its desired speed every 20 steps.

As explained in Sec.~\ref{sec:agentBehaviour}, an effect of the discretisation of the environment is the fact that diagonal movements generate a higher movement speed (as also discussed by the already cited~\cite{klupfel2003,DBLP:conf/acri/KretzKS08}, but also by~\cite{DBLP:conf/acri/SchultzF10}). In order to face this issue, this mechanism can be improved by considering these movements as a different kind of event during the extraction. With this strategy, each diagonal movement decreases the probability to move in the next steps according to the ratio $\Delta = \frac{0.4*\sqrt{2}-0.4}{0.4}$, which represents the relative difference between the space covered with a diagonal and linear movement. To discount diagonal movements, we introduced in the agents' state a parameter $diagPenalty$, initially set to 0, which is increased by $\Delta$ each diagonal movement. When $diagPenalty \geq 1$, the probability to move is decreased by adding in the urn of extraction one \emph{do not move} event or, in reference to Alg.~\ref{alg:HetUpdate}, by increasing of 1 unit the parameter $\beta$ after $updatePosition()$ invocation, decreasing $diagPenalty$ by 1.

This method is now consistent for reproducing different speeds for pedestrians in a discrete environment also considering the Moore neighbourhood structure. It must be noted, however, that if it is necessary to simulate very particular velocities (consider for instance a finer grained instantiation of a population characterised by a normal distribution of speed profiles), $Frac(\rho)$ is such that a large number of turns is needed to empty the urn, that is, to complete the movements to achieve an average speed equal to the desired one. This means that locally in time the actual speed of a pedestrian could differ in a relatively significant way from this value. To avoid this effect, during the life of each agent the fraction describing the probability is updated at each step and in several cases it will reach unreduced forms, with $GCD(\alpha,\beta)>1$. These situations can be exploited by splitting the urn into simpler sub-urns according to the $GCD$ value. For example, given a case with $Frac(\rho)=\frac{5}{11}$, after one movement the urn will be associated to $\frac{4}{10}$; since $GCD(4,10)=2$ the urn can be split into 2 sub-urns containing 2 \textit{move} and 3 \textit{do not move} events that will be consumed before restarting from the initial urn. The effect of this subdivision is to preserve a stochastic decision on the actual movement of the pedestrian but to avoid local excessive diversions from the desired speed.

\subsection{Modelling Stairs and Ramps}\label{sec:stairs}

Thanks to the above introduced mechanism for granting agents the possibility to have a desired speed in their state definition it is relatively simple to extend the model to be able to represent stairs and ramps.

Basically, we can represent areas in which the velocity of pedestrians is altered due to a slope by introducing an additional kind of marker that is perceived by agents passing in the related cells and that causes a modification of the desired speed. For simplicity we will now consider simple staircase and ramps characterised by a single bottom and top sections, that will be associated to markers indicating to the agent the entrance or exit from a particular area. Each marker $A_a$, associated to an area $A$, is characterised by two constants, $k_{a_1}, k_{a_2} \in \mathbb{R}$, respectively associated to the entrance to or exit from area $A$.

\begin{figure} [tbp]
\begin{center}
\includegraphics[width=.75\columnwidth]{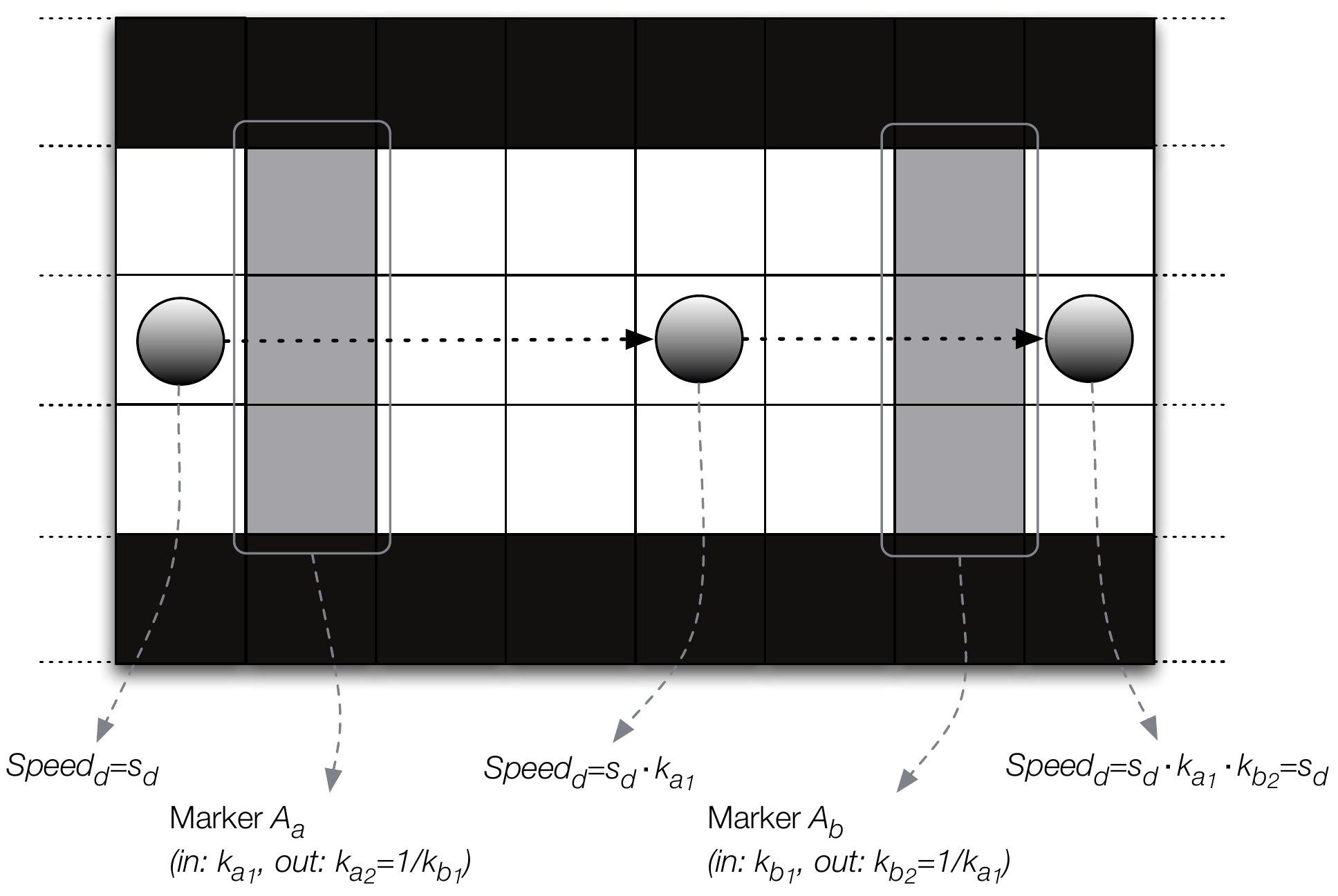}
\end{center}
\caption{A schema depicting speed modification markers and their potential application to model a small staircase.}
\label{fig:stairs}
\end{figure}

The state of an agent must be extended to include the area it is currently situated in; in this way, each agent perceiving the presence of a marker in the cell it has just moved in will (i) update the area it is currently situated in, (ii) update the desired speed according to the constant associated to the marker. An example of this procedure is shown in Figure~\ref{fig:stairs}: in this case, an agent is entering a staircase or ramp $A$ from the left side, delimited by the marker $A_a$, and exiting from the right side, delimited by the marker $A_b$. When first crossing the marker $A_a$ the agent will update its state to remember that the area it is currently situated in is $A$ and it will update its own speed by multiplying the current value by the constant $k_{a_1}$ (since it is entering the area). Then the agent moves to the right until it crosses marker $A_b$: the agent will now record that it is not situated in area $A$ anymore and it will update its desired speed by multiplying it by $k_{b_2}$.

\begin{figure} [tbp]
\begin{center}
\includegraphics[width=.6\columnwidth]{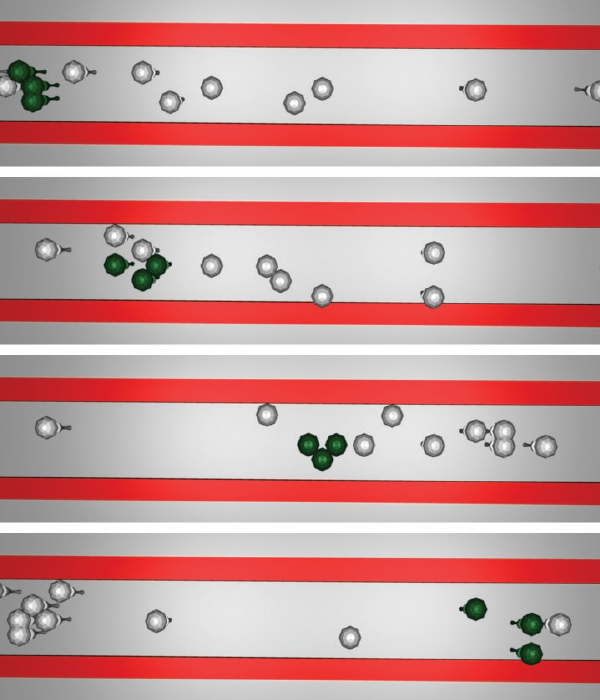}
\end{center}
\caption{Screenshots related to the movement of a simple group including pedestrians with heterogeneous speeds (in particular, a slowly moving person) throughout a corridor: the group cohesion mechanism is effective even when members are characterised by heterogenous desired speeds.}
\label{fig:screen}
\end{figure}

For a properly designed staircase or ramp, associated to an area $A$ and delimited by markers $A_a$ and $A_b$, we must enforce the constraints $k_{a_2}=1/k_{b_1}$ and $k_{b_2}=1/k_{a_1}$ in order to assure that each agent that completes a trajectory completely crossing the area (passing therefore both markers) returns to the initial desired speed.

\section{Simulation Results in Test Scenarios}

We carried out a simulation campaign to evaluate the effects of the introduction of this mechanism for the representation and management of heterogeneous speed profiles in the simulated population in reference scenarios in which empirical data can be found in the literature. We chose the scenarios in order to have a relatively good coverage of the relevant real world situations that one might encounter in a practical simulation project. In particular, we simulated and measured (i) mono directional flow of pedestrians in a corridor and a stair with varying density, (ii) bidirectional flow of pedestrians in a corridor with varying density, (iii) the merge of different flows of pedestrians in a T--junction. 

\begin{figure} [tbp]
\begin{center}
\includegraphics[width=.6\columnwidth]{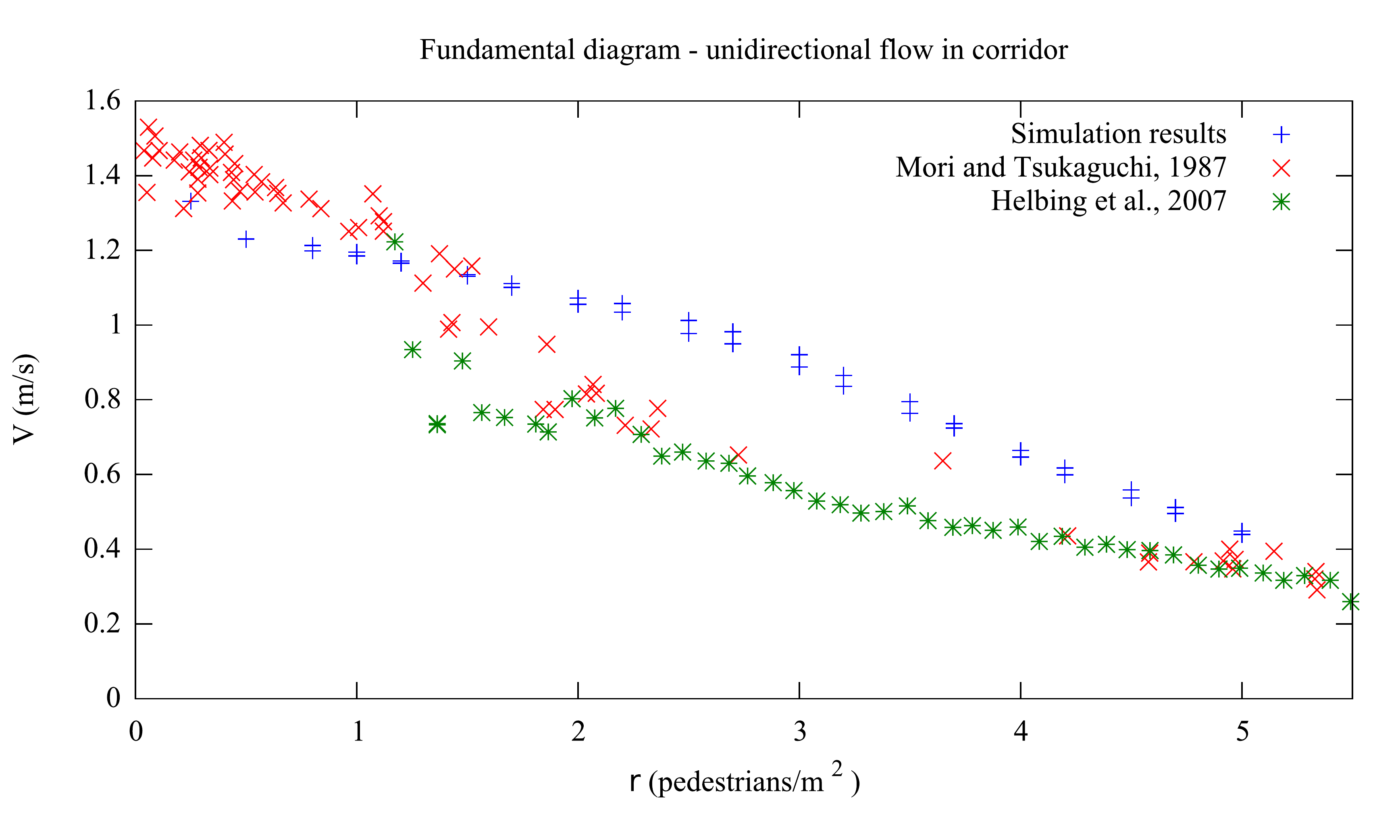}\\
\includegraphics[width=.6\columnwidth]{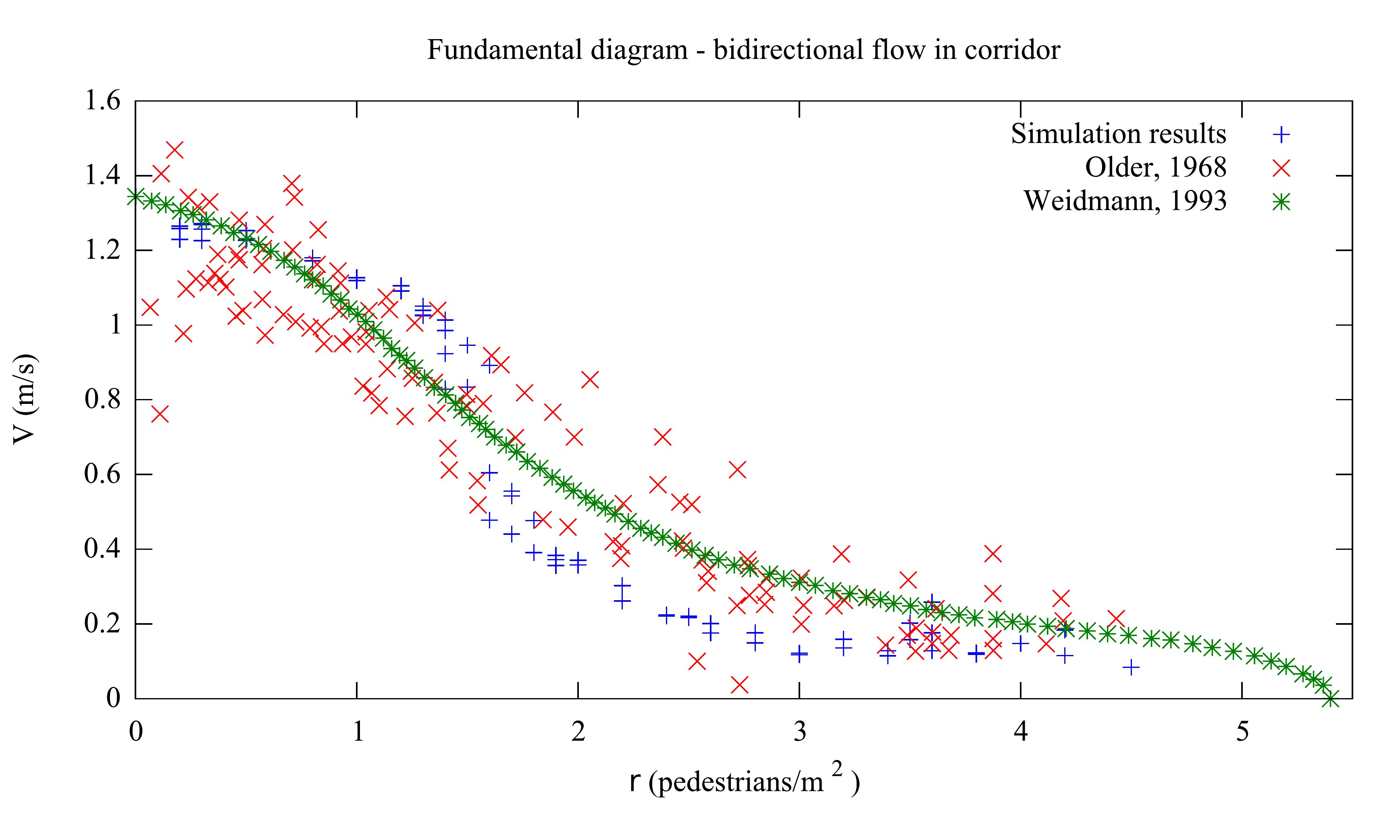}
\end{center}
\caption{Two fundamental diagrams respectively related to unidirectional flow (top diagram) and bidirectional flow (bottom one) in a corridor.}
\label{fig:fd}
\end{figure}

Before describing quantitative results, Figure~\ref{fig:screen} represents a qualitative result about the fact that, even with the introduction of the heterogeneous speed mechanism, the simple group cohesion mechanism is effective in avoiding excessive levels of dispersion. The depicted screenshots related to the movement of a simple group including pedestrians with heterogeneous speeds (in particular, a slowly moving person) throughout a corridor: despite the differences in individual speeds, the group preserves its cohesion, while being overtaken by other faster pedestrians. Such a qualitative result could be quantified by measuring the level of dispersion of simple groups in conditions of increasing density and changing environmental geometries, but we omit these results for sake of space and to focus on the effectiveness of the heterogeneous speed mechanism.

With reference to the simple linear scenarios (mono and bi directional flows in a corridor), we carried out a set of simulations with a growing number of pedestrians, and therefore density, measuring their average velocities to be compared with fundamental diagrams from the literature (see, e.g.,~\cite{DBLP:reference/complexity/SchadschneiderKKKRS09}). Since the model is now able to manage pedestrians with heterogenous walking speeds we configured the simulations to include pedestrians with desired speeds of $1.2$, $1.4$ and $1.6 m/s$ according to a gaussian--like distribution (in line with statistics on pedestrian behaviour that can be found, for instance, in~\cite{Willis2004}). The outcome of the simulation leads to results, shown in Figure~\ref{fig:fd}, that are relatively in tune with the empirical data from the literature: the velocities in case of unidirectional flow, velocities in densities between 2 and 4 pedestrians per square metre are slightly higher than empirical observations (in particular~\cite{MoriTsukaguchi-FD} and~\cite{PhysRevE.75.046109}); for bidirectional flow, velocities are very close to empirical results (in particular~\cite{older}) and also with design manuals (in particular~\cite{weidmann}). The combined results suggest that a more careful calibration of parameters involved in the rendering of social repulsion but not related to counterflow caused conflicts could improve the achieved results in the mono directional flow.

\begin{figure} [t]
\begin{center}
\includegraphics[width=.6\columnwidth]{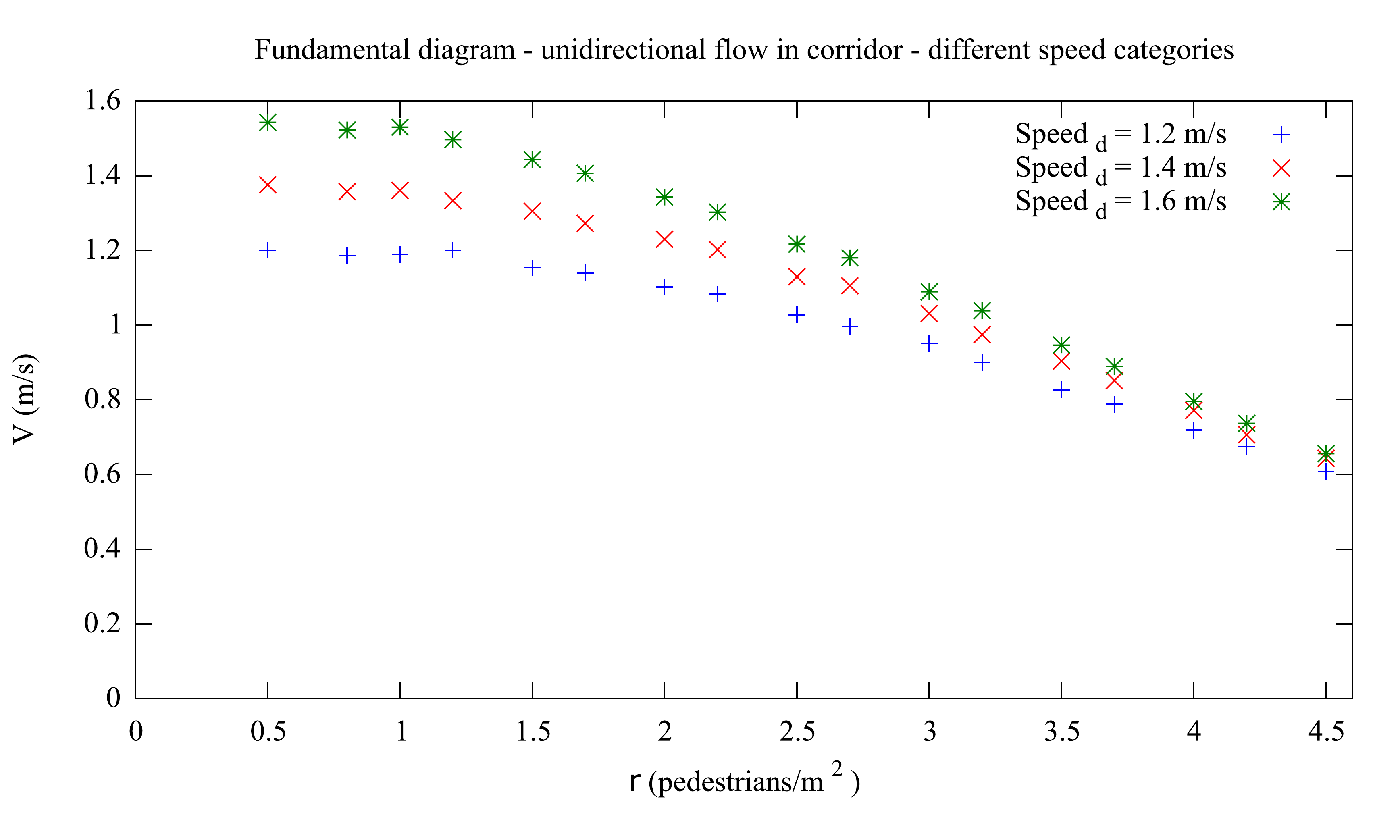}\\
\includegraphics[width=.6\columnwidth]{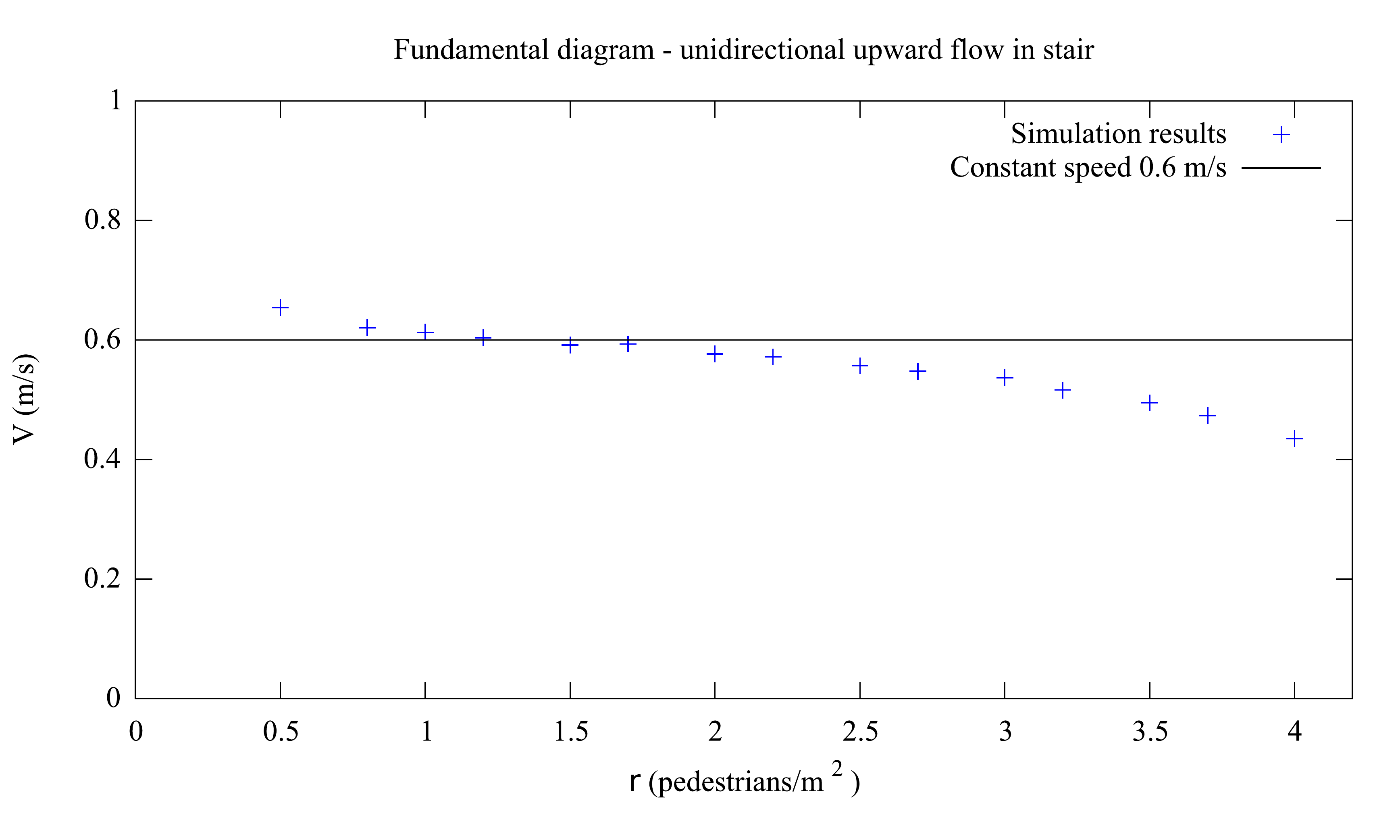}
\end{center}
\caption{Average walking speeds for different types of pedestrians (having different desired speeds)(top diagram), and average walking speed in a stair (bottom one), in different density conditions.}
\label{fig:het-speeds}
\end{figure}

We also evaluated the capability of the model to actually allow pedestrians to move at the actually desired speed and the conditions for this to happen: in Figure~\ref{fig:het-speeds}, on the left, we compare the average walking speed of pedestrians with different desired walking speeds in varying conditions. In free flow situations, pedestrians are generally able to move at a speed that is very close to the desired one. However, with the growth of the density, walking speeds decrease due to conflicts and this phenomenon is more significant for pedestrians having a higher desired speed, to the point that when reaching a certain density level they have essentially the same speed of pedestrians characterised by a lower walking speed. Figure~\ref{fig:het-speeds}, on the right, also shows the average speed in a stair section modelled as discussed in Section~\ref{sec:stairs} (pedestrians entering the stairs decrease their desired velocity by a factor of 2.33): also in this case, the actual average speed is close to the levels indicated by design manuals~\cite{Burghardt2013} and to available empirical data~\cite{DBLP:conf/case/DingLZC13}. We can conclude that the introduced mechanism is effectively able to reproduce heterogeneous walking speeds.

\begin{figure} [tbp]
\begin{center}
\includegraphics[width=.6\columnwidth]{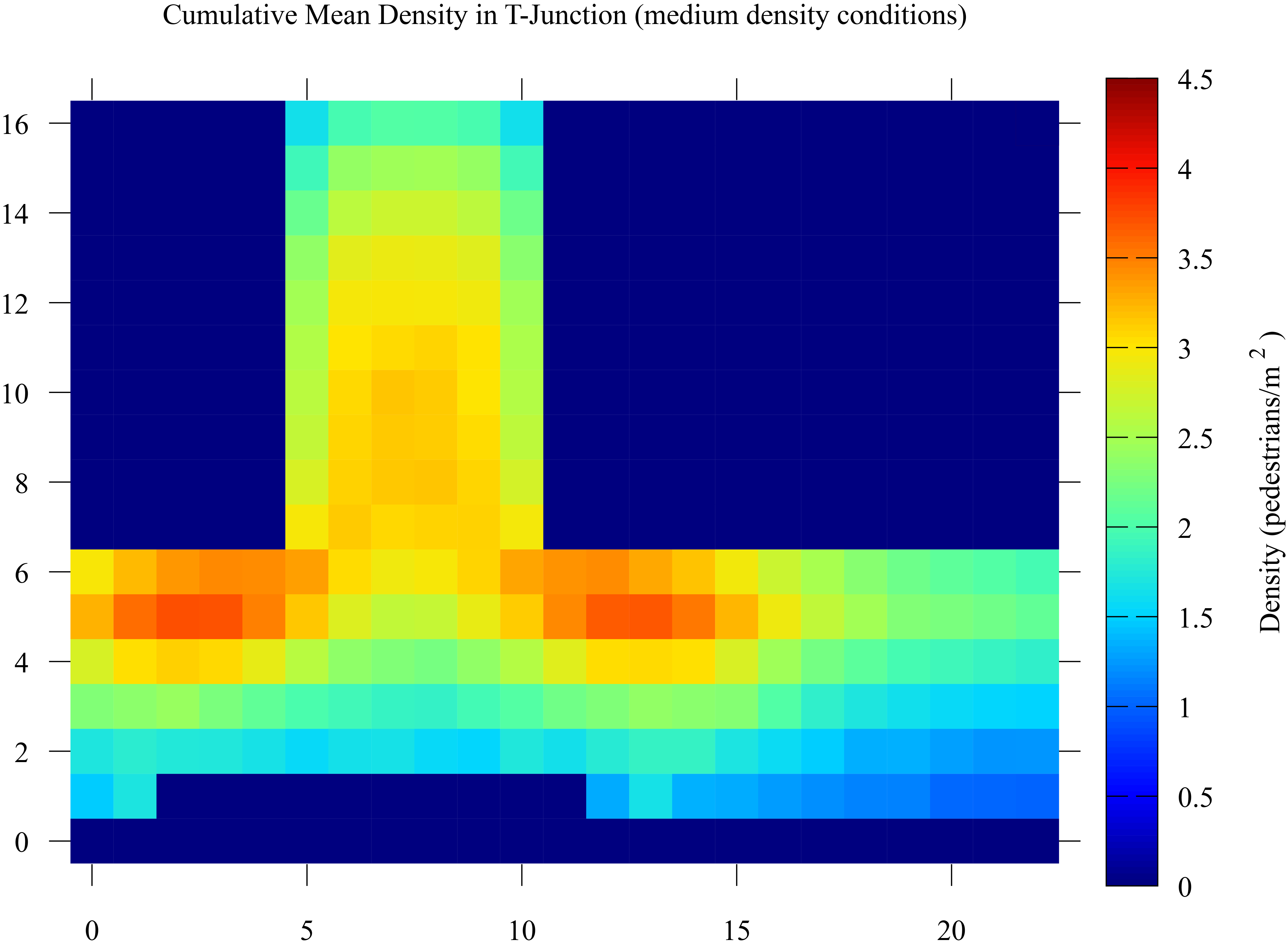}\\
\includegraphics[width=.6\columnwidth]{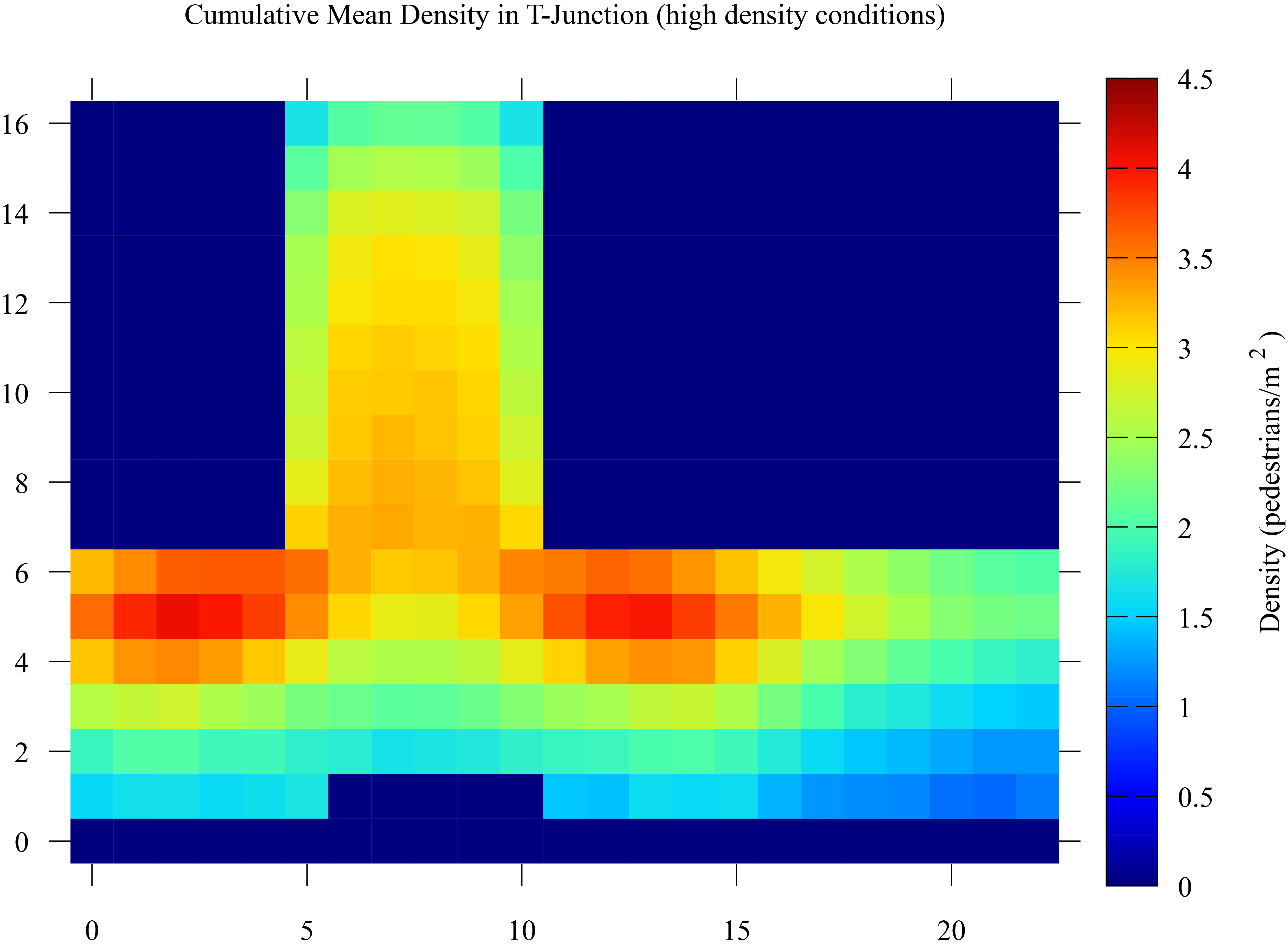}
\end{center}
\caption{Density diagram of pedestrian flow in the T-junction scenario, respectively in medium and high density situations.}
\label{fig:cmd}
\end{figure}

The final scenario is instead aimed at understanding if the defined model is able to reproduce patterns of space utilisation that are observable in reality, especially in environments including a corner, in which pedestrians are bound to make a turn. The T--junction scenario is characterised by two branches of a corridor that meet and form a unique stream. In this kind of situation it is not hard to reach high local densities, especially where the incoming flows meet to turn and merge in the outgoing corridor. The main result of this scenario is not represented by fundamental diagram data, but rather by an indication on how the pedestrians used the available space by moving in the environment throughout the simulation. We adopted a metric called \emph{cumulative mean density} (CMD)~\cite{Castle2011}, a measure associated to a given position of the environment indicating the average density perceived by pedestrians that passed through that point. It is quite straightforward to compute this value in a discrete approach like the one described in this work. As suggested above, we wanted to evaluate the capability of the model to reproduce patterns of spatial utilisation that are in good agreement with those resulting from the actual observations available in the literature~\cite{zhang2012}. Results shown in Figure~\ref{fig:cmd}, analysing situations of medium and high density, show qualitative phenomena (areas of low or high utilisation) that are in tune with available empirical data on this kind of scenario.

%\subsection{Bibliography}
%The TRB bibliography style is defined in the \verb1trb.bst1 file which should be
%in your document folder. A new command is specified, \verb1\trbcite{}1 which
%will print the authors and the number of the reference in the order in which it
%is supplied. The References section will be appended to the end of the document.

\section{Conclusions}
The paper has presented a technique for the introduction and management of heterogenous speeds in discrete pedestrian simulation models: the adopted model and mechanism has been formally described and results of its application to benchmark scenarios has been provided and discussed in relation to available data from the literature. Results show that the model and the introduced technique are effective in the reproduction of the target phenomenon and they represent a starting point for a more thorough application to the modelling of specific operative elements of environments, such as additional types of stairs and escalators. Additional experimentations and comparison with empirical data are necessary steps to consolidate this line of work that, on the other hand, will be also extended to consider dynamical aspects of group behaviour (formation and dissipation of groups, especially structured), but also coordination mechanisms like anticipation of counterflow conflicts, as discussed in~\cite{Suma2012248}.

\bibliographystyle{trb}
\bibliography{bibliography}

% End line numbering
\nolinenumbers
\end{document}